# Bayesian Space-time SIR modeling of Covid-19 in two US states during the 2020-2021 pandemic.

## Authors

Andrew B Lawson

Joanne Kim

Department of Public Health Sciences

Medical University if South Carolina

Charleston SC

USA

## Abstract

This paper describes the Bayesian SIR modeling of the 3 waves of Covid-19 in two contrasting US states during 2020-2021. A variety of models are evaluated at the county level for goodness-of-fit and an assessment of confounding predictors is also made. It is found that models with three deprivation predictors and neighborhood effects are important. In addition, the work index from Google mobility was also found to provide an increased explanation of the transmission dynamics.

## Keywords

Space-time;  Bayesian  Covid19  modeling;  SIR  Google mobility deprivation South Carolina  New Jersey  goodness of fit  propagator models

# Introduction

The Covid-19 pandemic is now the focus of interest for many researchers as it persists into 2022. There is now a large body of research around the pandemic (see eg. https://www.who.int/emergencies/diseases/novel-coronavirus-2019/global-research-on-novel-coronavirus-2019-ncov ,https://www.paho.org/journal/en/special-issues/scientific-papers-and-resources-covid-19, https://www.thelancet.com/coronavirus; https://connect.medrxiv.org/relate/content/181 ) and more specifically focused on modeling of the spread of Covid in different locales (https://covid19scenariomodelinghub.org/).

Amongst modeling efforts there is a range of approaches depending on purpose. While the pandemic progresses dynamically, it is clear that the spatial spread of the disease impacts different sub-areas within countries. Many papers examine aggregate time series of cases and/or deaths without reference to the spatial context of infection (IHME2021; Bjornstadt et al., 2020). While this can be useful at a given scale it is less useful when considering resource allocation decisions to different communities.

There is also a divide between those who use deterministic compartment models, usually based on differential equations ( usually ODEs) and those who use a statistical modeling basis. In the former case, the statistical properties of the dynamic infection process are not addressed (see e.g., Xiang et al, 2021). While error estimates are often cited for these methods they are not based on statistical assumptions. Of course,statistical assumptions can be embedded in this process either by assuming that the ODEs are the underlying mean level within a statistical model or assuming that the ODEs are stochastic. Often this is not specified however.

On the other hand, for statistical modeling approaches a dichotomy often arises between purely descriptive and mechanistic models. First, descriptive models have been used to provide a summary of the Covid19 behavior ( see e.g. Carroll and Prentice, 2021). These models essentially use general random effect components to mimic the space-time behavior of the epidemic process, and these models can be effective in terms of goodness of fit. The alternative is to assume a mechanistic model for the epidemic behavior which acknowledges the transmission process and allows for both mechanistic and descriptive components. This latter approach provides for the estimation of parameters that arise in the context of deterministic ODE models but also manages to account for extra random variation. This has the advantage that prediction can be based on transmission parameters while also allowing for evaluation of statistical property of estimators such as mean estimates and credible/confidence or prediction intervals.

Additionally, a Bayesian parametric approach to modeling epidemic spread is advantageous as it provides for exact model-based inference and the ability to specify and test prior distributions for parameters in their relevance to the modeling process (Lawson and Song, 2010; Lawson and Kim, 2021; Sartorius et al, 2021).

In this paper, we adopt a Bayesian modeling paradigm focused on a mechanistic statistical model which allows the inclusion of both fixed (predictor) effects and random effects. Unlike previous modeling approaches (Lawson and Kim, 2021), the aim of the modeling is to assess the best model for multiple waves of Covid19 at the county level within the US. In addition, the focus is on two different states within the US (South Carolina and New Jersey) chosen for different policy directives with respect to behavioral interventions (non-pharmaceutical interventions: NPIs), and to assess both a set of deprivation confounders and the role of mobility data is the dynamic of the epidemic as it developed. We have considered the period of March 6$^{th}$, 2020 until February 21$^{st}$, 2021, a period of some 353 days which includes 3 waves of the pandemic.  Figure 1 displays the county maps of both SC and NJ.

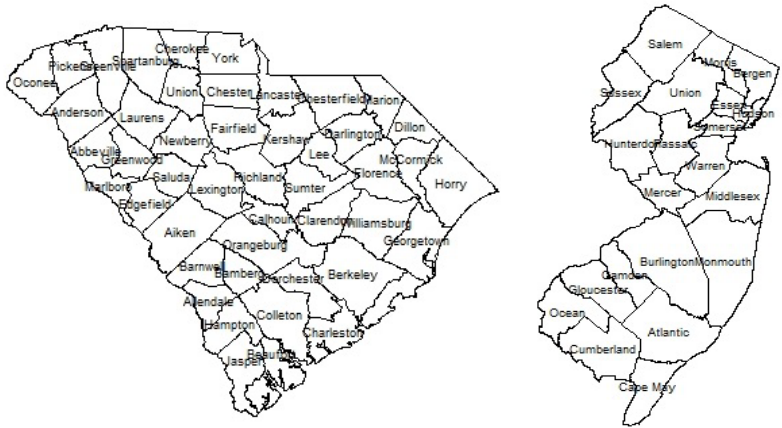

*Figure 1 South Carolina (L) and New Jersey (R) county maps*

**Data: cases and deaths**

Our data consists of daily recorded cases and deaths for each county for each of South Carolina (46 counties) and New Jersey (21 counties). Cases were reported from Health departments and deaths from the National Center for Health Statistics. These data are available from a range of sources and we used the NYT GitHub repository raw files for this purpose (https://github.com/nytimes/covid-19-data ). The data used consists of daily case and death counts for the period of 353 days. Based on the recording biases induced by weekends and holidays, we have also aggregated the case event data and averaged it over a 3-day period (3-day average).

**Data: predictors**

In addition, we have also obtained county-specific deprivation measures: % of population below the poverty line, % black ethnicity and finally a recent measure of deprivation from the American Community Survey: the Multidimensional Deprivation Index (MDI) from 2017 (https://www.census.gov/library/publications/2019/acs/acs-40.html ).  Finally, we also examined mobility data to help assess the effects of lockdown restrictions in different regions. To this end, we obtained the daily Google mobility data for the counties of SC and NJ for residential and work indices (https://www.google.com/covid19/mobility/ ). Other Google indices were examined but were not used as they contained missing data or did not reflect common mobility behavior across both states. For example, the transit index is not relevant in SC as few people use public transport, and other indices are colinear with the indices used.

Google mobility data has some drawbacks in that it relies on the location finder being activated on Google on a device. This suggests that it is likely to be unascertained. Because of this, we also explored alternative mobility sources. We did obtain SC data on traffic flows for major highways, which are available on a county basis (https://www.scdot.org/travel/travel-trafficdata.aspx ), but extensive missingness, for certain counties and time periods, prohibited their use and so this was not pursued.

To assess the impact of lockdowns, which mainly happened in 2020, we focused our analysis of Google mobility data on the period of 301 days from March 6th, 2020 to December 31st, 2020.

**Data Models**

Our mechanistic model for case counts follows from the susceptible-Infected-removed (SIR) model of Morton and Finkenstadt (2005) as extended to the spatial domain by Lawson and Song (2010) and extended for Covid19 modeling by Lawson and Kim (2021). Note that the exposed compartment within an SEIR model is not observable in these data and would be estimated without testable data. We wish to avoid extra assumptions where at all possible so this is not admitted in our models. The basic Poisson count data model is assumed for the case count $y_{ij}$ in the *i* th area and *j* th time period, for *m* areas and *J* time periods, with susceptible population $S_{ij}$:

$y_{ij} \sim Pois(\mu_{ij})$ with $\mu_{ij} = S_{ij}.f(y_{i,j-1},........)$, so that the susceptible population is modulated by a function of previous counts and other related effects. Critiques of Poisson data models often focus on the mean-variance equality and resort to negative binomial data models to allow overdispersion (e. g. Lloyd-Smith, 2007). However it is straightforward to demonstrate that over-dispersed Poisson models result from the inclusion of random effect terms in mean specification and as these are specified straightforwardly in a

Bayesian setting then there is little need to reject the Poisson data model. In a later section the ability of Poisson models with time-dependent random effects to capture Covid-19 variation is demonstrated clearly (see e.g. Figure 4).

In what follows we define

$$log(\mu_{ij}) = log(S_{ij}) + log(f(y_{i,j-1},........))$$
$$= log(S_{ij}) + p^r_{i,j-1}$$

where $p^r_{i,j-1}$ is a propagator and can be specified with a variety of terms thought to influence the transmission over space and time. These terms could include lagged counts of disease (in the same unit or neighboring units), fixed covariates, and also random terms.

Note that under a SIR model an accounting equation must be specified which updates the susceptible populations dynamically. Hence it is assumed that

$$S_{ij} = S_{i,j-1} - I_{i,j-1} - R_{i,j-1}$$

where $I_{i,j-1}$ is the true infective load, and $R_{i,j-1}$ is the removal from the population. The true infective load could consist of observed counts plus an asymptomatic component (unobserved). In our models we assume that $I_{i,j-1} = y_{i,j-1} + y^{as}_{i,j-1}$ where $y^{as}_{i,j-1}$ is assumed to be a simple proportion of the observed case load. We assume here $y^{as}_{i,j-1} = \lambda y_{i,j-1}$ where $\lambda$ can be pre-specified from information on asymptomatic rates. Over the pandemic, different estimates have been proposed for $\lambda$ ranging from 0.25 (20%) up to 1.0 (50%) ( see e.g. Ma et al. 2021). In a later section we report the results of using different $\lambda$ values in the model specification.

We used an above Poisson count model representation for the daily case counts data.

In the case of 3-day (3D) averaged case counts, which are continuous, we assume a different data model. In fact for the 3D average $y^{3D}_{ij}$ a log normal model with mean $\mu_{ij}$ is assumed with the following form

$$y^{3D}_{ij} = exp(y^*_{ij});$$
$$y^*_{ij} \sim N(\mu_{ij}, \tau^{-1})$$

The specification of the mean function can follow as for propagators for the Poisson model as these are on the log scale. Note that the addition of a variance parameter in the data model allows more flexibility at that level.

The issue of vaccination strategy in response to the pandemic has been considered within the modeling framework. However, the end date of this example is in early 2021 and given that vaccination role-out for the general population had not been established by that date, coupled with the fact that most of the period studied did not have any vaccination coverage, led us to exclude this in our modeling. A further complication is that age distributions are used to role-out vaccination and we did not have access to population age sub-strata.

**Propagator models:**

A number of possible propagator models have been considered for both SC and NJ case counts.

Our base model from which a variety of extensions are made is model 1:

$$p^r_{i,j-1} = \alpha_0 + \alpha_1 log(y_{i,j-1}) + v_i + u_i \quad (1)$$

In this model there is a constant intercept and a regression parameter for the log of the previous case count within the same region (county). To allow for random confounding noise we include a uncorrelated and spatially correlated random effect ($v_i, u_i$). These have the following prior distributions:

$$v_i \sim N(0, \tau_v^{-1})$$
$$u_i | \{u_k\}_{k \neq i} \sim N(\bar{u}_{\delta_i}, \tau_u^{-1}/n_{\delta_i})$$

where $\delta_i$ is a neighborhood of the *i* th area, $n_{\delta_i}$ is the number of neighboring areas if the *i* th area, and $\bar{u}_{\delta_i}$ is the mean of the *u* values in the neighborhood of the *i* th area. The prior distribution for $u_i$ is known as intrinsic conditional autoregressive (ICAR). The combination of the two effects is known as a convolution model. Note that the ICAR model depends on the adjacencies of areas so that neighborhoods can be defined.

Base variants of the above model have been examined for the county level data on Covid-19.

$$p^r_{i,j-1} = \alpha_0 + \alpha_1 log(y_{i,j-1}) + v_i \quad (2)$$
$$p^r_{i,j-1} = \alpha_0 + \alpha_1 log(y_{i,j-1}) + \alpha_2 log(\sum_{k \in \delta_i} y_{k,j-1}) + v_i + u_i \quad (3)$$
$$p^r_{i,j-1} = \alpha_0 + \alpha_1 log(y_{i,j-1}) + \alpha_2 log(\sum_{k \in \delta_i} y_{k,j-1}) + v_i \quad (4)$$

In addition these base variants have been augmented with predictors representing deprivation and mobility. Initially only deprivation predictors (% under the poverty line, % black ethnicity, Multidimensional Deprivation Index 2017) were examines either as fixed predictor effects, or selected with Gibbs variable selection (GVS: Dellaportas et al., 2002; O'Hara, R. and M. J. Sillanpää, 2009). GVS allows the sampling of a range of different model combinations using entry parameters and leads to estimates of the inclusion probability for different predictors.

The following represent the final model variants fitted to the SC and NJ county level data. An asymptomatic rate of 20% was assumed in these models. Models 4A – 5B have an added spatial term ($\sum_{k \in \delta_i} y_{k,j-1}$)) which is the sum of case counts in neighboring areas at the previous time period. This is included as disease spread between areas is likely and dependence on this spread could play a role in current transmission.

1) $p^r_{i,j-1} = \alpha_0 + \alpha_1 \log(y_{i,j-1}) + v_i + u_i$

2A) $p^r_{i,j-1} = \alpha_0 + \alpha_1 \log(y_{i,j-1}) + v_i$

2B) $p^r_{i,j-1} = \alpha_0 + \alpha_1 \log(y_{i,j-1}) + v_i$   plus fixed deprivation predictors

3A) $p^r_{i,j-1} = \alpha_0 + \alpha_1 \log(y_{i,j-1}) + v_i$   GVS on deprivation predictors

3B) $p^r_{i,j-1} = \alpha_0 + \alpha_1 \log(y_{i,j-1}) + v_i + u_i$   GVS on deprivation predictors

4A) $p^r_{i,j-1} = \alpha_0 + \alpha_1 \log(y_{i,j-1}) + \alpha_2 \log(\sum_{k \in \delta_i} y_{k,j-1}) + v_i + u_i$   GVS on deprivation predictors

4B) $p^r_{i,j-1} = \alpha_0 + \alpha_1 \log(y_{i,j-1}) + \alpha_2 \log(\sum_{k \in \delta_i} y_{k,j-1}) + v_i$   GVS on deprivation predictors

4C) $p^r_{i,j-1} = \alpha_0 + \alpha_1 \log(y_{i,j-1}) + \alpha_2 \log(\sum_{k \in \delta_i} y_{k,j-1}) + v_i$

5A) $p^r_{i,j-1} = \alpha_0 + \alpha_1 \log(y_{i,j-1}) + \alpha_2 \log(\sum_{k \in \delta_i} y_{k,j-1}) + v_i$   plus fixed deprivation predictors

5B) $p^r_{i,j-1} = \alpha_0 + \alpha_1 \log(y_{i,j-1}) + \alpha_2 \log(\sum_{k \in \delta_i} y_{k,j-1}) + v_i + u_i$   plus fixed deprivation predictors

**Model Fitting**

The above models were all fitted using MCMC software on R. Specifically we used Nimble to fit the models and report the goodness of fit and parameter estimates. The posterior mean parameter estimates are reported here. Convergence was checked using the Geweke diagnostic in CODA. The mean deviance and WAIC were collected. The WAIC is widely used measure of fit adjusted for the degree of parameterization (Gelman et al, 2014) .

**South Carolina models**

Table 1 reports the overall goodness-of-fit metrics for the above models using mean deviance and WAIC. For models with GVS the three posterior mean inclusion probabilities are reported (models 3A – 4B).

*Table 1 South Carolina initial fitted models*

| Model | Deviance(mean) | WAIC |
|---|---|---|
| 1) | 434242.8 | 163911 |
| 2A) | 434240.2 | 163072 |
| 2B) | 434240.5 | 163377.8 |
| 3A) | 434242.1 | 163758.7 |
| 3B) | 434243.1 | 163450.4 |
| 4A) | 420571.1 | 163395.9 |
| 4B) | 420568.9 | 161418.1 |
| 4C) | 420569.0 | 161790.6 |
| 5A) | 420568.3 | 160748.2 |
| 5B) | 420570.4 | 161229.7 |

The best model based on WAIC is model 5A which includes the neighborhood effect, an uncorrelated random effect and fixed predictors. The UH model is best overall (WAIC: *160748.2).* Note that models with ICAR effects are not favored, although the spatial neighborhood effect is. It is notable that all models with neighborhood propagator have lowest WAIC. For GVS models the threshold for inclusion in the final model is usually assumed to be 0.5 (Barbieri and Berger, 2004). The fixed predictor model (5A) yielded the parameter estimates in Table 2. It is notable that none of the quantile estimates cross zero which suggested the parameters are well estimated.

*Table 2 South Carolina model 5A parameter estimates*

| Parameter | Posterior mean | Posterior 2.5 %tile | Posterior 97.5 %tile |
|---|---|---|---|
| $\alpha_0$ | -10.083 | -10.112 | -10.052 |
| $\alpha_1$ | 0.191 | 0.190 | 0.192 |
| $\alpha_2$ | 0.078 | 0.0767 | 0.079 |
| $\beta_1$ | 0.084 | 0.081 | 0.086 |
| $\beta_2$ | -1.646 | -1.799 | -1.502 |
| $\beta_3$ | -0.829 | -0.974 | -0.661 |

Figure 2 displays the uncorrelated heterogeneity surface for the best fitted model.

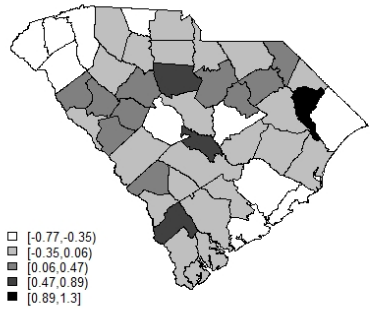

*Figure 2 Posterior mean uncorrelated effect for model 5A*

Figure 3 displays the case profiles with posterior mean risk estimates from the best fitting model for Beaufort and Charleston counties. It is noticeable that while the model tracks well the overall variation in risk, there remains considerable random noise. The Appendix includes examples of modelled profiles of risk for selections of counties both for SC and NJ. The first of these is model 5A for the case data from SC.

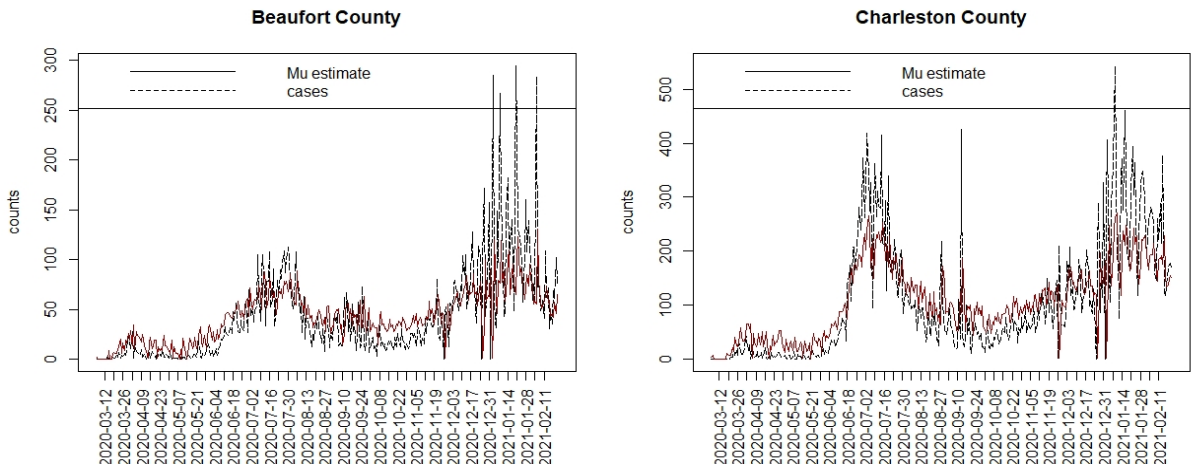

*Figure 3 Case count and posterior mean risk estimate for Beaufort (L) and Charleston (R) county*

**Some SC model variants**

While the best daily case model for SC provides a reasonable fit, it is clear that random variation is evident and changes in the noise are time varying. A model allowing for these changes would provide a better fit overall.

1) **Allowing for variable heterogeneities between areas and over time**

To address this random variation, we have assumed that the uncorrelated heterogeneity can vary over space and time as well. We have assumed a prior distribution of $v_{ij} \sim N(0, \tau_v)$ an uncorrelated prior distribution, but we also looked at time variation in the variability of $v_{ij} \sim N(0, \tau_{vj})$, which allows the precision of the term to vary: with $\tau_{vj} \sim Ga(0.5, 0.1)$. In the first case we get a mean deviance of 78505 and WAIC of 88654.6 In the second case we get 76528 for the mean deviance and 86586.7 for the WAIC. It is clear that time varying random effect with variable precision provides a considerably reduced WAIC. Figure 4 displays the results of fitting the time varying precision model for two counties in SC. It is clear that this model closely models the variation. However, given that a considerable extent of variation is attributed to random noise, this model may be considered less useful for predictive purposes, and possibly over-parameterized.

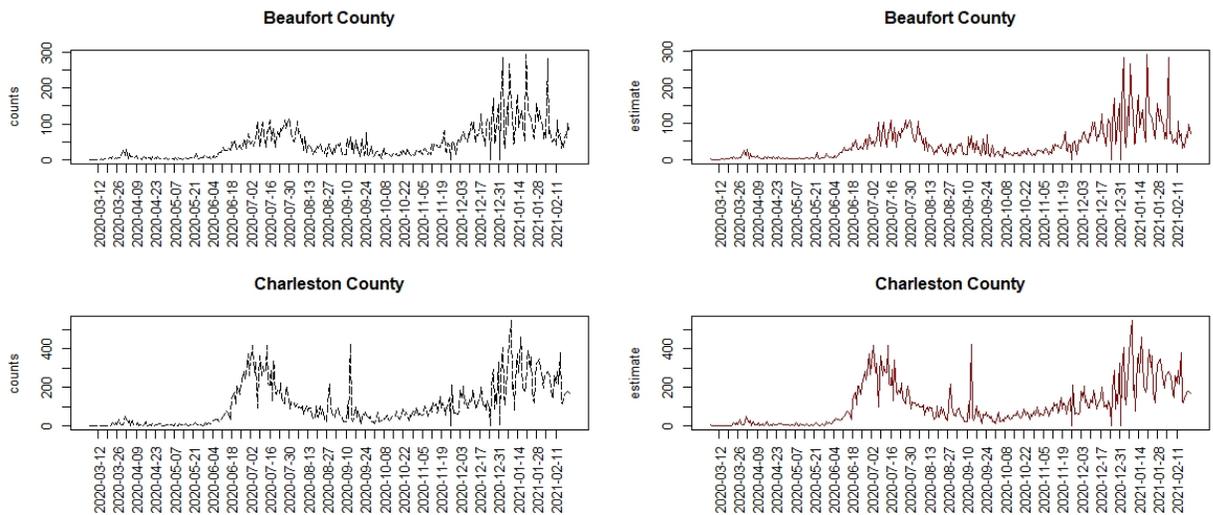

*Figure 4 Comparison of two SC counties under a model with time varying random effect and time varying precision. Left: case count; right: posterior mean estimated risk.*

### 2) Different asymptomatic rates

We examined the effect of changing the asymptomatic rate on goodness of fit. Initial models were fitted with 20% asymptomatic rate. We then looked at 25, 50 and also 75%. We examined these new rates with model 5A as above.

*Table 3 Asymptomatic rate results : South Carolina*

| Asymptomatic rate | Mean deviance | WAIC |
|---|---|---|
| 20% | 420568.3 | 160748.2 |
| 25% | 412569.3 | 161935.3 |
| 50% | 425328.3 | 163277.5 |
| 75% | 459951.0 | 167270.0 |

It appears that the lowest rate (20%) has the lowest WAIC although 25% has lowest deviance. This seems to align with the current reviews of asymptomatic spread discussed above.

### 3) SC models for 3-day average smoothed Data

The 3-day (3D) averaged data was also examined using a log-normal data model of the form:

$$y_{ij}^{sm} \sim LN(\mu_{ij}, \tau_y^{-1})$$
$$\log(\mu_{ij}) = \log(S_{ij}) + p_{i,j-1}^r$$

A variety of propagator models were also examined as follows:

1) $p_{i,j-1}^r = \alpha_0 + \alpha_1 \log(y_{i,j-1}) + u_i$
2) $p_{i,j-1}^r = \alpha_0 + \alpha_1 \log(y_{i,j-1}) + v_i + u_i$
3) $p_{i,j-1}^r = \alpha_0 + \alpha_1 \log(y_{i,j-1}) + u_i + v_{ij}$

As the data model also has a precision assigned then it is possible to also consider model 2) but with $\tau_{yj}$ time dependent (model 4)

Table 4 3D data results: South Carolina

| MODEL | WAIC | Comments |
|---|---|---|
| 1 | 176455.9 | |
| 2 | 176497.8 | |
| 3 | 102893.3 | $v_{ij}$ time dependent |
| 4 | 95699.6 | $\tau_{yj}$ time dependent |
| 5 | 103848.9 | Only $v_i$ but with fixed predictors |
| 6 | 103826.9 | As model 5 plus Neighborhood lagged effect |

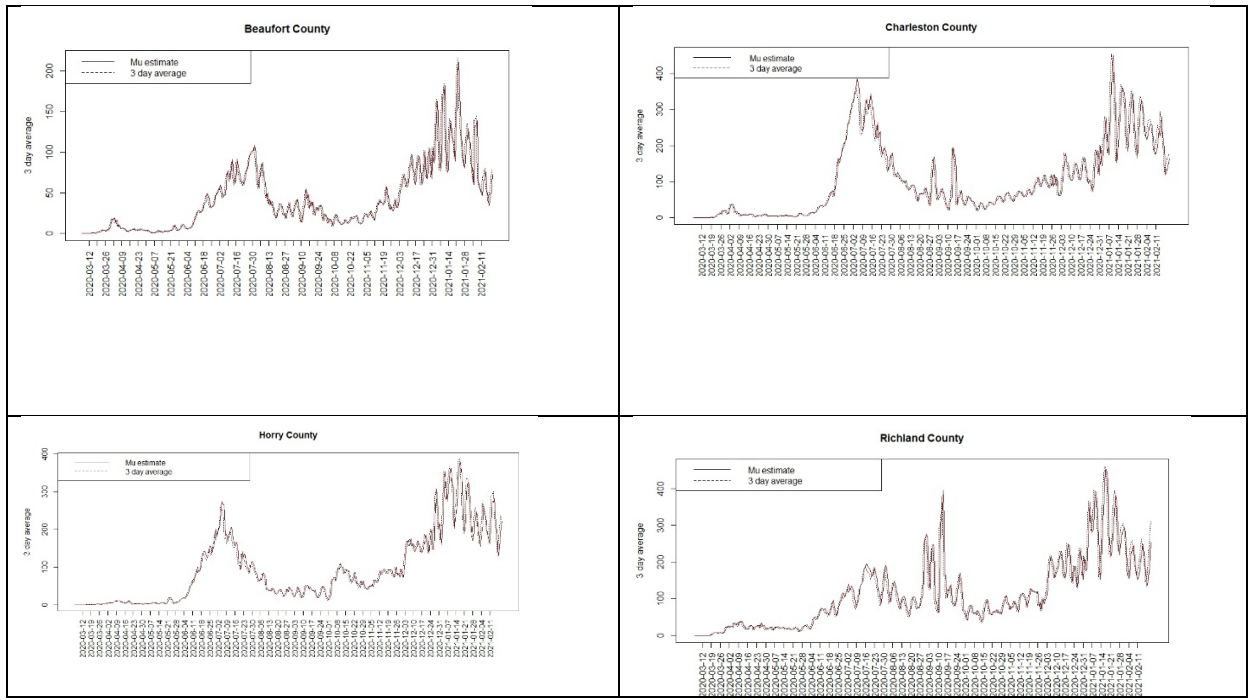

*Figure 5 Comparison of four SC counites with fitted model 4*

Model 4 appears to fit well with time dependent precision. There is no neighborhood effect or spatial CAR component and no predictors included. However, as discussed above, this could be over parameterized, and so we would like to provide better explanation (and hence predictive power) by using predictors and other terms which can provide explanation better without resort to random components.

Hence model 5 and model 6 include a fixed set of the 3 deprivation predictors. Note that we found previously that GVS did not provide better model fits based on WAIC and so we did not include these in this comparison:

model (5): $p^r_{i,j-1} = \alpha_0 + \alpha_1 log(y_{i,j-1}) + v_i$ and fixed deprivation predictors which has *WAIC= 103848.9*

model (6): $p^r_{i,j-1} = \alpha_0 + \alpha_1 log(y_{i,j-1}) + \alpha_2 \sum_{k \in \delta_i} log(y_{k,j-1}) + v_i$ and fixed deprivation predictors

which was the best fitting for the Poisson daily models. This has WAIC = 103826.9

Hence the neighborhood effect does reduce the WAIC over model 5. However, it is notable that when you allow the variability to be time dependent then the WAIC is much more reduced. The lowest WAIC is for time dependent UH or time dependent data model precision. Without these effects the best models are those with predictor effects.

For the predictor effect models, model 6 resulted in the following posterior results.

*Table 5 3D smoothed data: parameter estimates for model 6*

| Parameter | Mean | SD | 2.5 Percentile | 97.5 Percentile |
|---|---|---|---|---|
| $\alpha_0$ | -10.941 | 0.097 | -11.107 | -10.750 |
| $\alpha_1$ | 0.898 | 0.003 | 0.892 | 0.904 |
| $\alpha_2$ | 5.49595x 10-6 | 1.11352x 10-6 | 3.25060x 10-06 | 7.60140 x10-6 |
| $\beta_1$ | 0.101 | 0.015 | 0.070 | 0.123 |
| $\beta_2$ | 0.454 | 0.527 | -0.263 | 1.882 |
| $\beta_3$ | 0.114 | 0.630 | -1.022 | 1.573 |
| $\tau_y$ | 0.946 | 0.011 | 0.924 | 0.968 |

In summary, for model 6 the propagator parameters, including the neighborhood term, are all well estimated. The deprivation predictor parameters are not well estimated except for $\beta_1$ which is associated with the % under the poverty line, although the ethnicity predictor was marginal. The MIDI17 predictor was not well estimated.

Figure 6 displays the posterior mean county uncorrelated effect under model 6. There is some suggestion that the upper midland area is under-estimated but in general the map suggest a random patterning which supports a good model fit.

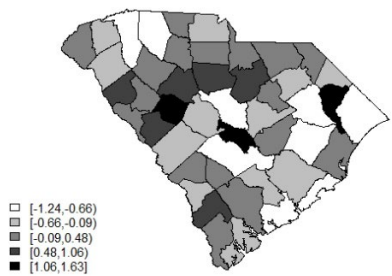

*Figure 6 Posterior mean uncorrelated random effect for model 6*

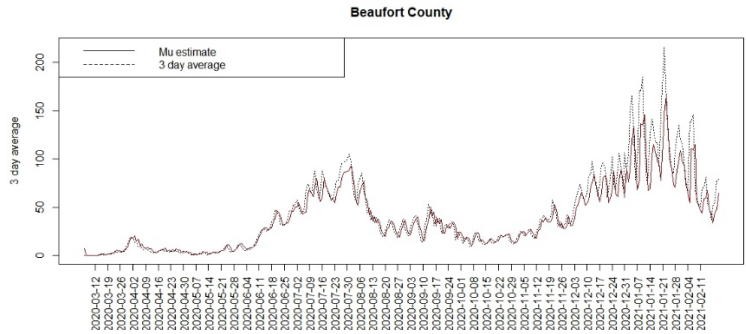

*Figure 7 Count and posterior mean risk for Beaufort county under model 6*

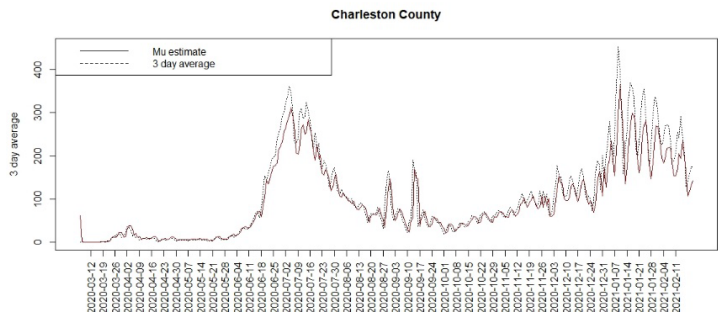

*Figure 8 Count and posterior mean risk for Charleston county under model 6*

Figure 7 and 8 display the model 6 profiles of case counts and estimated risk. The model it appears to be reasonably good overall with the lowest WAIC, besides the time dependent random effect models. The Appendix includes a selection of six county profiles of risk under the model 6 for the 3D average data.

**New Jersey models**

The state of New Jersey is a northern US state with a much larger population with greater urbanization, by contrast to the largely rural and lower population density in South Carolina. It is of interest to examine how models perform when applied to this contrasting state. For the same period of 353 days (03/06/2020 – 02/21/2021), we have fitted a range of models equivalent to those for South Carolina. Figure 9 displays the case count and 3D average for selected county in NJ (Hunterdon). It is clear a number of waves had occurred over the same period as SC, but there is also a major spike in case load in the NJ data, that appears in all county profiles, due to a delay in recording cases over the new year period of 2020. This spike clearly can influence a daily case model. Nevertheless, for comparisons purposes, we report here the results of fitting a daily case count Poison models as per SC example.

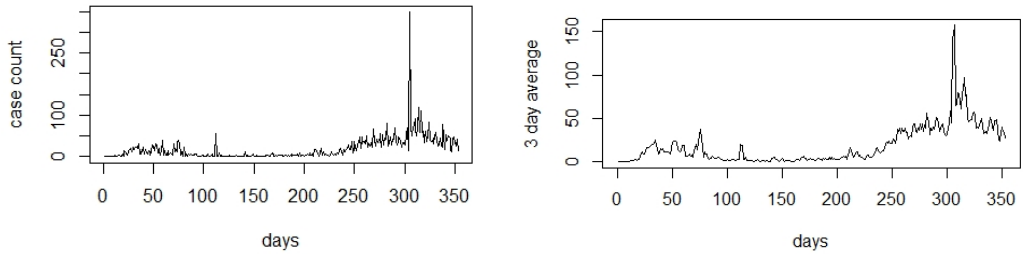

*Figure 9 Hunterdon county case count (left) and 3D average (right) : March 6th 2020 to February 21st 2021.*

Table 6 describes the results of the Poisson daily count model fits in terms of goodness-of-fit. The models were the same as in the basic models for South Carolina but added for model 6 the time dependent heterogeneity term with prior distribution

$v_{ij} \sim N(0, \tau_v^{-1})$     6 a)  or

$v_{ij} \sim N(0, \tau_{vj}^{-1})$  with  $\tau_{vj} \sim Ga(0.5, 0.1)$   6 b)

*Table 6 New Jersey initial count data daily Poisson models*

| Model | WAIC | Deviance (mean) | Comments |
|---|---|---|---|
| 1) | 413255.5 | 405174.7 | |
| 2A) | 413264.5 | 405174.1 | |
| 2B) | 413055.1 | 405173.9 | |
| 3A) | 411675.5 | 404211.7 | |
| 3B) | 413528.1 | 405175.1 | |
| 4A) | 361015.5 | 353949.5 | |
| 4B) | 360672.5 | 353948.6 | |
| 4C) | 360920.6 | 353949.0 | |
| 5 A) | 360745.2 | 353948.9 | $v_i$ +Fixed predictors |
| 5B) | 360423.0 | 353251.7 | 5A) + ICAR |
| 6A) | 52189.3 | 47097.9 | Fixed predictors + $v_{ij}$ |
| 6B) | 52121.3 | 46893.3 | Fixed predictors + $\tau_{vj}$ |

It is notable that time dependent random effect models (6 a,b) have lowest WAIC as in the SC case. However they have a large random component and so potentially unattractive for any predictive purpose. It is clear that model 5B followed by 4B have the lowest WAIC values. These models both include predictors within their formulation.

These models include a lagged neighborhood effect (4B) with uncorrelated heterogeneity, with 5B an extension with fixed predictors and inclusion of a ICAR effect.

$$p_{i,j-1}^r = \alpha_0 + \alpha_1 log(y_{i,j-1}) + \alpha_2 log(\sum_{k \in \delta_i} y_{k,j-1}) + v_i \quad \text{plus GVS predictors (4B)}$$

$$p_{i,j-1}^r = \alpha_0 + \alpha_1 log(y_{i,j-1}) + \alpha_2 log(\sum_{k \in \delta_i} y_{k,j-1}) + v_i + u_i \quad \text{plus fixed predictors (5B)}$$

As we are interested in the effect of deprivation predictors on transmission then we will consider model 5B further. Note that both a lagged neighborhood effect and an ICAR effect are included in 5B which suggests that there is residual spatial correlation structure in the data, unlike the SC example. Table 7 displays the posterior mean parameter estimates for this model.

*Table 7 New Jersey daily count data model 5B posterior mean parameter estimates*

| Parameter | Mean | SD | 2.5 Percentile | 97.5 Percentile |
|---|---|---|---|---|
| $\alpha_0$ | -10.318 | 0.036 | -10.369 | -10.247 |
| $\alpha_1$ | 0.255 | 0.001 | 0.254 | 0.256 |
| $\alpha_2$ | 0.187 | 0.001 | 0.185 | 0.188 |
| $\beta_1$ | 0.115 | 0.007 | 0.102 | 0.128 |
| $\beta_2$ | 0.053 | 0.003 | 0.049 | 0.058 |
| $\beta_3$ | -3.847 | 0.584 | -4.945 | -2.909 |

All the transmission parameters and regression parameters are well estimated.

Both % under poverty and % Black are positively associated with risk while the multidimensional deprivation index is negatively related as the lower the deprivation the less risk. This contrasts with the SC result where % poverty is positive but % black is negative. Of course a different best model arises in the SC case and so it is to be expected that a different direction in estimation is likely. The best SC model does not include an ICAR term. The Appendix includes a selection of six county risk profiles for model 5B for the NJ case data.

### New Jersey 3D averaged data models

For comparison with SC we have used the 3 day (3D) averaged data in model scenarios.

*Table 8 New Jersey Smoothed data models*

| Model | WAIC | Deviance (mean) | Comments |
|---|---|---|---|
| 1 | 64598.9 | 78370.2 | |
| 2 | 64600.4 | 78374.41 | |

| | 3 | 63545.0 | 84802.4 | $u_i + v_{ij}$ time dependent |
|---|---|---|---|---|
| | 4 | 52563.7 | 84888.4 | $u_i + v_{ij}$, $\tau_{yj}$ time dependent |
| | 5 | 81580.1 | 47459.6 | $v_i$ + fixed predictors |
| | 6 | 61881.6 | 81126.5 | As model 5 plus neighborhood lagged effect |

For the smoothed NJ data the lowest WAIC is found for the model 4 which includes a spatial convolution and time dependent precision on the outcome model. If the time dependent predictor models are excluded (due to their emphasis on the random component), then model 6 with neighborhood effect, uncorrelated random term and fixed predictors has the lowest WAIC. It is notable from Table 9 that the regression estimates for the deprivation predictors from model 6 are not well estimated, and it can be noted that a simpler model without deprivation predictors had a WAIC value close to model 6 but not lower. The Appendix includes a selection of six county profiles for model 6 for the NJ 3D average data.

| Parameter | Mean | SD | 2.5 Percentile | 97.5 Percentile |
|---|---|---|---|---|
| $\alpha_0$ | -12.467 | 0.159 | -12.737 | -12.148 |
| $\alpha_1$ | 0.925 | 0.005 | 0.916 | 0.934 |
| $\alpha_2$ | 9.424349e-08 | 1.233779e-07 | -1.508541e-07 | 3.317263e-07 |
| $\beta_1$ | 0.038 | 0.041 | -0.039 | 0.115 |
| $\beta_2$ | -0.031 | 0.016 | -0.056 | 0.009 |
| $\beta_3$ | 0.074 | 0.625 | -1.120 | 15 |

Table 9 New Jersey 3D data: Posterior mean parameter estimates for model 6

**Mobility data: South Carolina and New Jersey**

Mobility of population is a major factor in the potential for disease spread and so changes in mobility would be a useful indicator which may influence transmission rates. For example, neighborhood effects could be modified by work-at-home orders and so measures that relate to this change in pattern could add to risk explanation. We have examined two potential sources of mobility data for our examples. Initially, in the case of South Carolina we explored the use of the Department of Transport's (SCDOT) records of traffic counts. These are available for a selection major traffic intersections and highways in the state. The data is partially incomplete in that certain counties have no traffic record and also time profiles for many others are only partially available. In addition, county aggregate data is based on a subset of road intersections and there is some doubt as to how representative these would be. While we considered imputation to provide more complete statewide data, it was considered that the incompleteness could significantly impact the analysis and could lead to considerable uncertainty in model appropriateness. For these reasons we decided to focus on analysis based on Google mobility data, which is available for both SC and NJ.

**Google mobility data**

Since early February 2020, Google has recorded indices of locational activity for many locations around the globe, with the intention to provide data for Covid-19 researchers(https://www.google.com/covid19/mobility/ ). The data is based on location switch set on the google phone app and depends crucially on whether people with mobile phones have used this setting. To that extent then these data are under-ascertained, in that only phone users and those who choose to switch on a location access in Google are recorded. In that this is a subset of the population it represents a sample of behaviors during the pandemic.

Google provides daily mobility data in the form of 6 indices representing different aspects of movement: work, transit, residential, retail, grocery and parks. The indices are measured relative to the location at a base date in early February 2020. Across SC and NJ there are a number of periods where data is missing for a number of these indices. The most complete indices for both areas are work and transit. However transit is less useful in SC as few people use public transport. Hence here we confine our report to the work profiles for each state. Figures 10 and 11 display the loess-smoothed profiles for the work index for both states (right panel). It is clear that there is a relation between mobility and transmission but it is not constant over time. Hence to accommodate these temporal changes in the relation, we have considered change points throughout the period studied (March 6$^{th}$ 2020 to December 31$^{st}$ 2020). To facilitate this analysis we have sought to determine inflection points where decisions on lockdowns or other NPIs are made and subsequently relaxed.

In the case of South Carolina, there are 4 periods when decisions changed. The first period of 18 days was from March 13$^{th}$ ( state of emergency declared) until March 31$^{st}$ (full lockdown), followed by partial lifting periods of 20 and 13 days. The final lifting of mandatory restrictions happened on May 12$^{th}$ . We have assumed that a four periods of 18, 20, 13, and 250 days are relevant as change points.

In the case of New Jersey there were three periods with an extended initial lockdown

The first was a minor lockdown between March 9th and 21st (16 days since the March 6th), followed by a restricted lockdown of 80 days until June 9th. Finally the lifted lockdown was lifted for 205 days during our monitoring period. Hence we assume that 16, 80, 205 are reasonable changepoints for the New Jersey example.

We examined three model variants assuming a Poisson daily count data model and 20% asymptomatic rate. A base model (model 1) consisting of the fixed deprivation predictors, an uncorrelated random effect and lagged case count and neighborhood case count combined with the mobility index:

$$p_{i,j-1}^r = \alpha_0 + \alpha_1 log(y_{i,j-1}) + \alpha_2 log(\sum_{k \in \delta_i} y_{k,j-1}) + v_i + x_i^T \beta + \eta_{[tch[j]]} \cdot w_{ij}$$

where $\eta_{[tch[j]]}$ is the parameter pertaining to the time indexed regression period ($tch[j]$) and $w_{ij}$ is the work index. Model 2 and 3 are variants of model 1 with mobility removed (model 2), and with mobility kept in but the neighborhood effect removed (model 3). Table 10 displays the goodness of fit results for these models. It is clear that in both cases model 1, the base model, is favored. Table 11 and 12 displays the posterior estimates of the parameters for the transmission effects and the mobility predictors.

Table 10 WAIC goodness of fit for mobility models

| Model | South Carolina | New Jersey |
|---|---|---|
| 1 | 268029.3 | 137691.2 |
| 2 | 283867.0 | 138540.7 |
| 3 | 278683.4 | 155839.7 |

| SC parameter | Mean | 2.5 percentile | 97.5 percentile |
|---|---|---|---|
| $\beta_1$ | 0.0708 | 0.0686 | 0.0733 |
| $\beta_2$ | -0.7005 | -0.9933 | -0.3255 |
| $\beta_3$ | 0.6327 | 0.1462 | 1.0833 |
| $\eta_{[tch[1]]}$ | 0.0466 | 0.0419 | 0.0516 |
| $\eta_{[tch[2]]}$ | 0.0219 | 0.0210 | 0.0229 |
| $\eta_{[tch[3]]}$ | 0.02377 | 0.0226 | 0.0249 |
| $\eta_{[tch[4]]}$ | -0.0164 | -0.0169 | -0.0159 |

Table 11 Posterior mean parameter estimates for the SC mobility model 1

*Table 12 Posterior mean parameter estimates for the NJ mobility model 1*

| NJ parameter | Mean | 2.5 percentile | 97.5 percentile |
|---|---|---|---|
| $\beta_1$ | 0.0605 | 0.0569 | 0.0646 |
| $\beta_2$ | 0.0979 | 0.0965 | 0.0994 |
| $\beta_3$ | -15.022 | -15.341 | -14.525 |
| $\eta_{[tch[1]]}$ | -0.0061 | -0.0077 | -0.0045 |
| $\eta_{[tch[2]]}$ | -0.0004 | -0.0050 | -0.0038 |
| $\eta_{[tch[3]]}$ | -0.0083 | -0.0089 | -0.0078 |

In these tables the change point parameters are denoted $\eta_{[tch[j]]}$ and in all cases these are well estimated, attesting to the relevance of the mobility work index in explaining variations in case transmission. It is also notable that a different pattern appears between SC and NJ for the deprivation variables ($\beta_s$), and while all are significant they appear to have different direction and degree of association.

Figures 10 and 11 display the model results for a selected county from each area for the mobility base model.

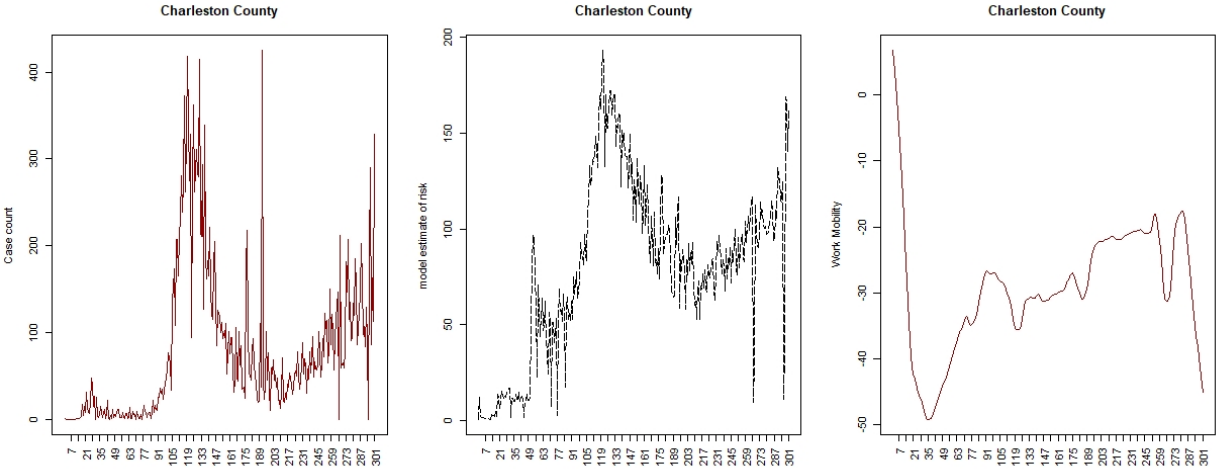

*Figure 10 SC: Charleston county case count, modelled risk, work mobility profile*

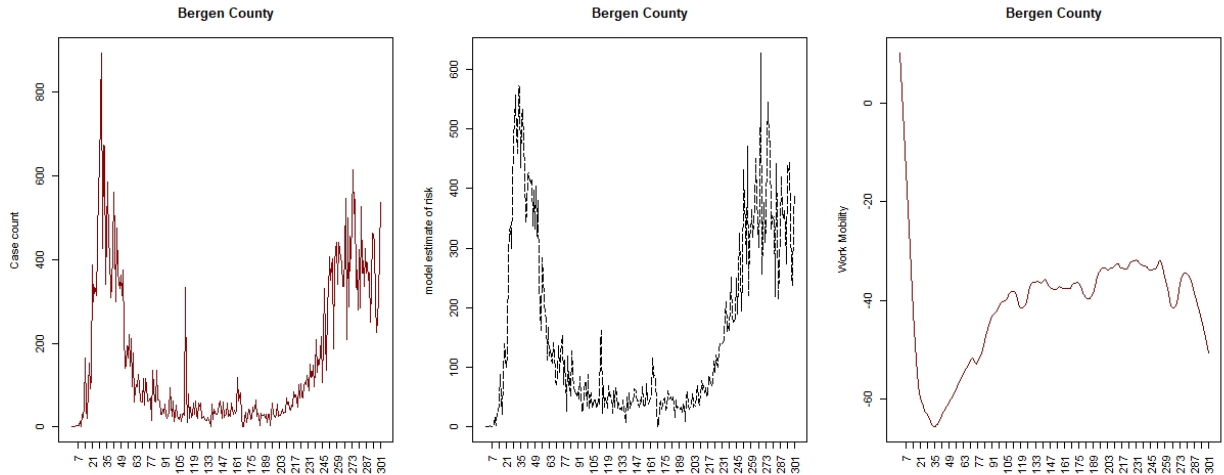

*Figure 11 NJ: Bergen county case count, estimated risk and work mobility profile.*

Comparison of the mobility results suggest the following: 1) mobility increases earlier and more steeply in SC than NJ , largely due to the early lifting of the lockdowns; 2) the mobility time periods are all significant in both states; 3) all SC deprivation predictors are significant also : % poverty (+); ethnicity (-); MIDI17 (+);  4) all NJ predictors significant also : % poverty (+); ethnicity (+); MIDI17 (-); 5) the mobility parameters for SC are +,+,+,- whereas the parameters for NJ are  -,-,-.  This latter effect mirrors the differences in case rate variation over time. It is also notable that % under the poverty line is significant and positive associated with increased transmission in both states.

The Appendix includes a selection of six county estimated risk profiles for both SC and NJ mobility predictor model 1. It is notable that while each area is best fit by mobility model 1, the fit is clearly better for NJ than for SC counties. This is apparent in the Appendix figures and it does suggest that work mobility was a better descriptor of the transmission dynamic in NJ than in SC, possibly due to the adherence to a longer lockdown in the latter state.

**Comparison between South Carolina and New Jersey**

It is notable that for the basic Poisson SIR models for both states then neighborhood lagged effects, three fixed predictors and uncorrelated heterogeneity `with 20% asymptomatic rate provide the best model. In both cases, all three predictors (% under the poverty line, % black ethnicity and MDI) are well estimated. For the NJ data an extra spatial correlation term is also needed. This suggest that there is some residual correlation unaccounted for by the neighborhood lagged effect. For the 3 day smoothed data, the time dependent models provide close data fits but lead to a large random component. When these models are discounted, then for SC, the best models include a neighborhood lagged effect and fixed predictors with % under the poverty line as the only well estimated predictor. For NJ a similar situation arises but no predictors are significant for the smoothed data. Overall then it is noticeable that for smoothed data, which may be preferred for prediction, the predictors are much less important in providing a good model fit.

In general, inclusion of mobility data does provide extra explanation of the transmission variation and it is to be noted that the deprivation parameter estimation strength can change when mobility is included. Poverty remains across all analyses as a significant predictor of transmission risk however.

**Conclusions**

A number of conclusions can be drawn from this study. First the ability of the SIR models to explain the spatio-temporal variation in Covid-19 risk has been demonstrated. The importance in these models of lagged neighborhood effects is significant and suggests that simple time series models do not adequately explain the spatial variation. The spatial context of disease spread is therefore important in such modeling.

Second, it is always possible to add temporal random effect terms in these models to gain close approximation to the case time series in various areas. Hence overdispersion modeling is straightforward, and there is no necessity to resort to negative binomial data models for this purpose, given the Bayesian context. However, it also suggests that over-parameterization can occur and from a prediction standpoint is not potentially useful.

For smoothed data models (3D average) the predictor effects are less important, and for SC the poverty predictor only remained important, while for NJ none of the deprivation predictors remained well estimated. As smoothed data models are often used for prediction, then it is worth noting that prediction could be made reasonably accurately without the inclusion of such predictors, especially in the NJ case.

While the mobility data examined was limited to the Google work index, it was demonstrated that it had a significant impact on the transmission using the change points selected, and also on the estimation of deprivation predictors. For the time periods chosen all mobility parameters were significant and all derivation predictors were also. It seems that the effect of different patterns of lockdowns affected mobility and in turn affected transmission. The interaction between mobility and deprivation seems to vary however, with only the poverty predictor remaining unchanged and positively associated with risk. The ethnicity and MDI17 predictors switched direction between SC and NJ and hence were not consistently associated, albeit well estimated.

In terms of future directions, we would like to examine vaccination strategy in relation to transmission dynamics when sub-strata of populations are available and for later time periods during 2021. In addition, the issue of effectiveness of lockdowns in different areas would be another avenue for further work.

# Appendix

Included here are figures demonstrating the model fitting capabilities of the best models found for each of three scenarios: case counts, 3D averages, and case counts with mobility data.

## Daily Case counts

SC: model 5A

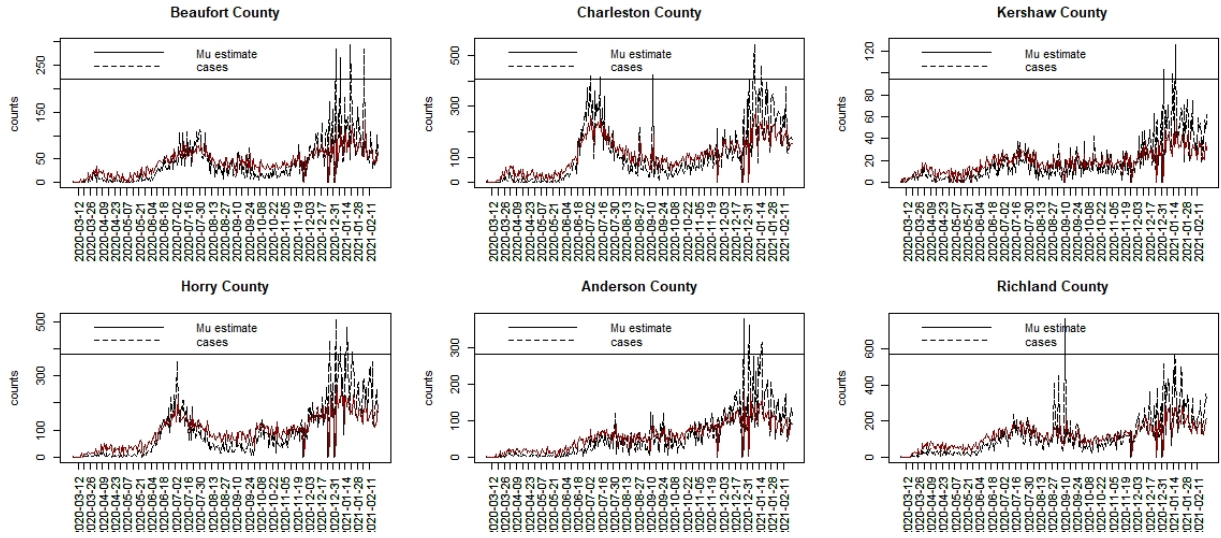

NJ: model 5B

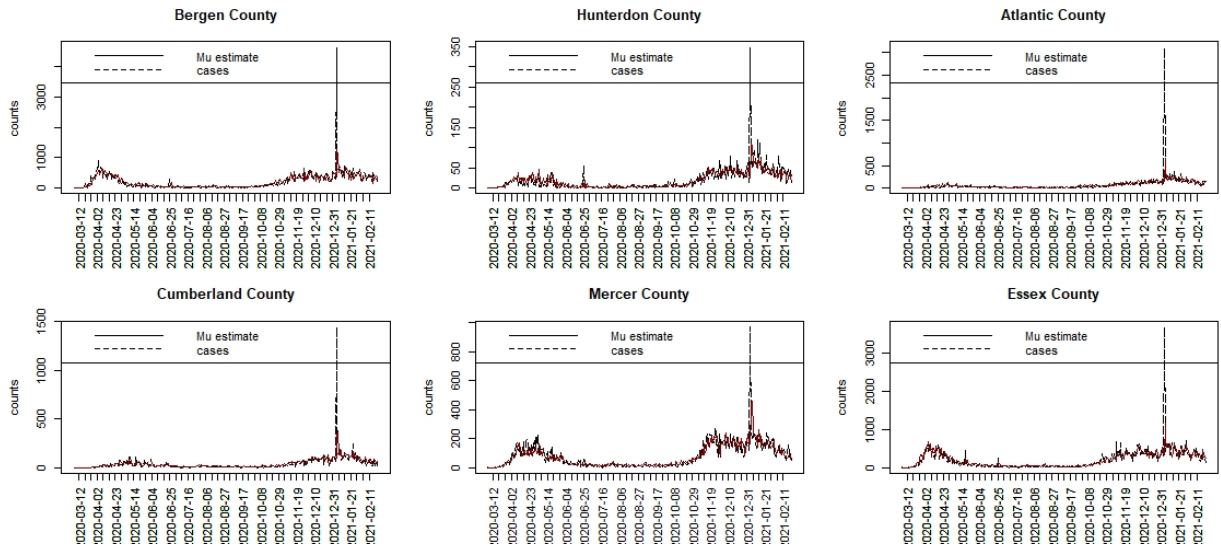

## 3Day averaged data

SC: model 6

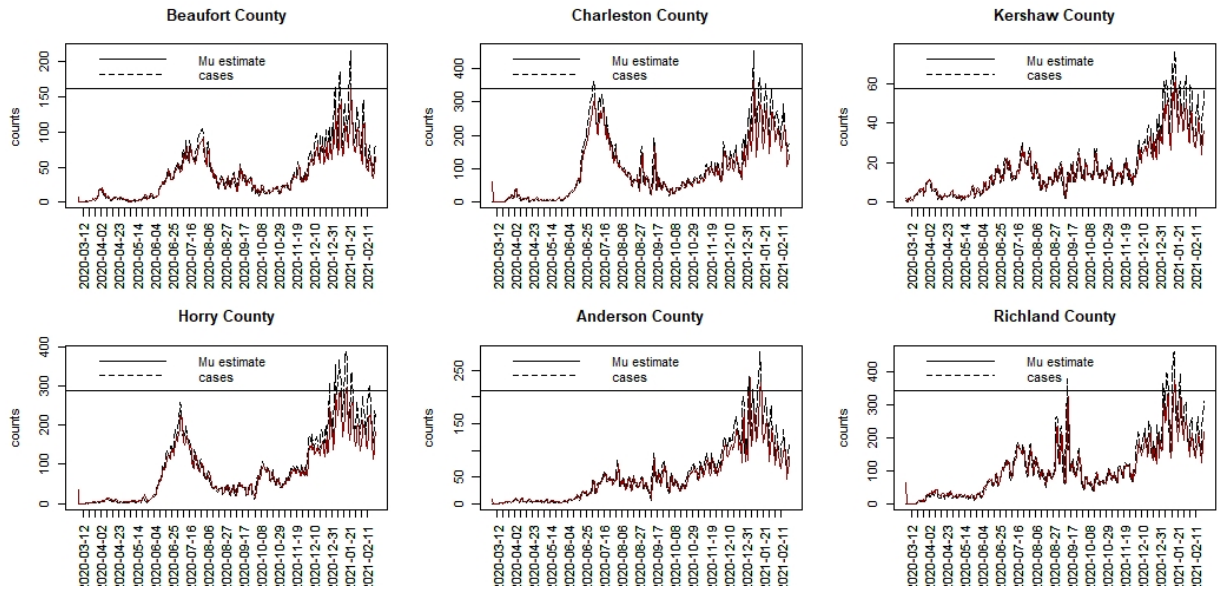

NJ: model 6

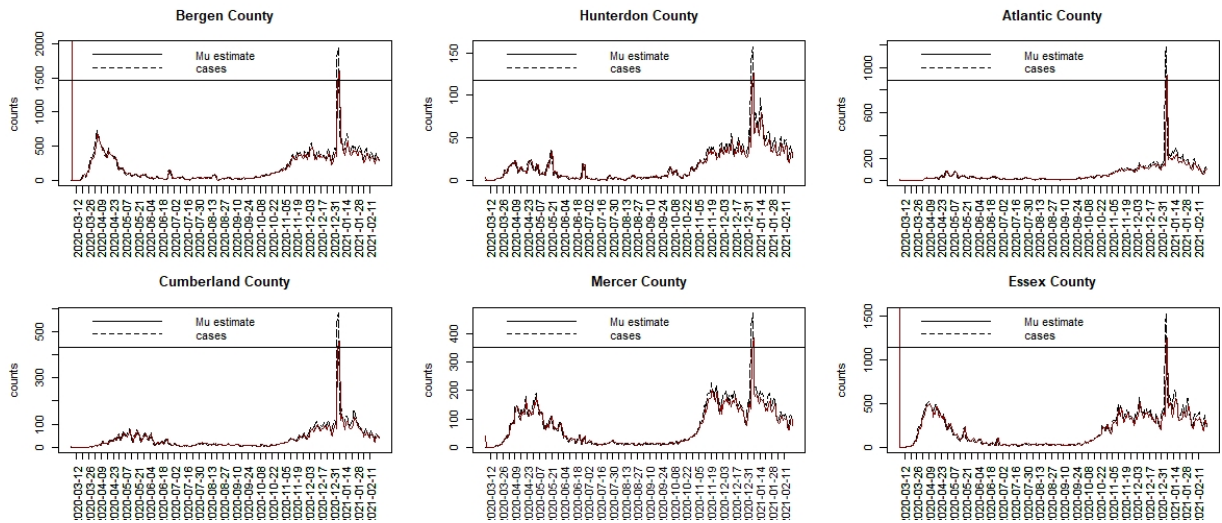

**Mobility data: work index**

SC: model 1

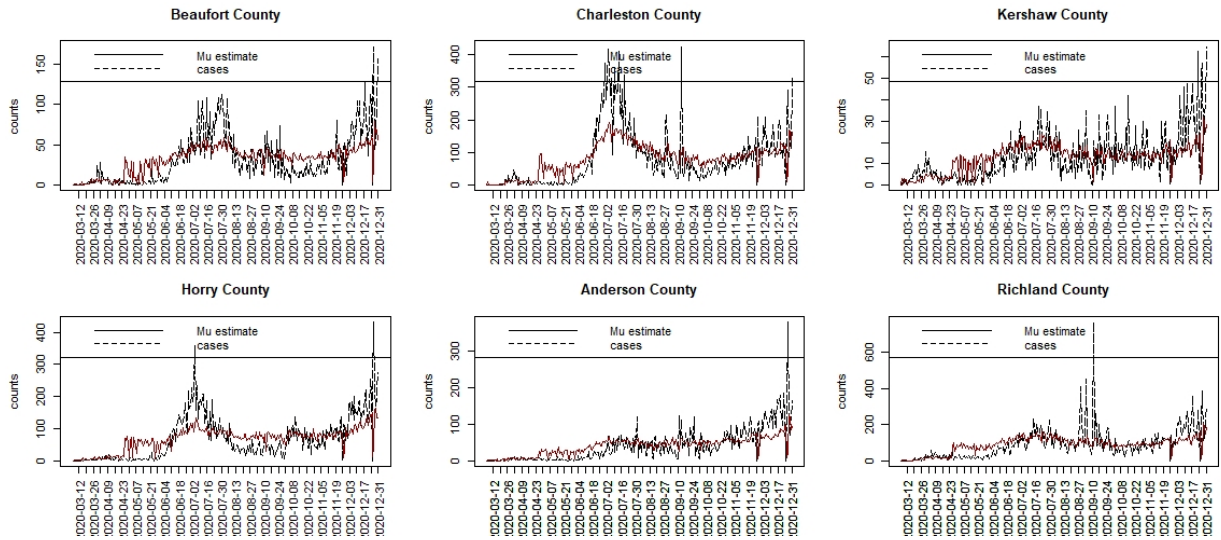

NJ: model 1

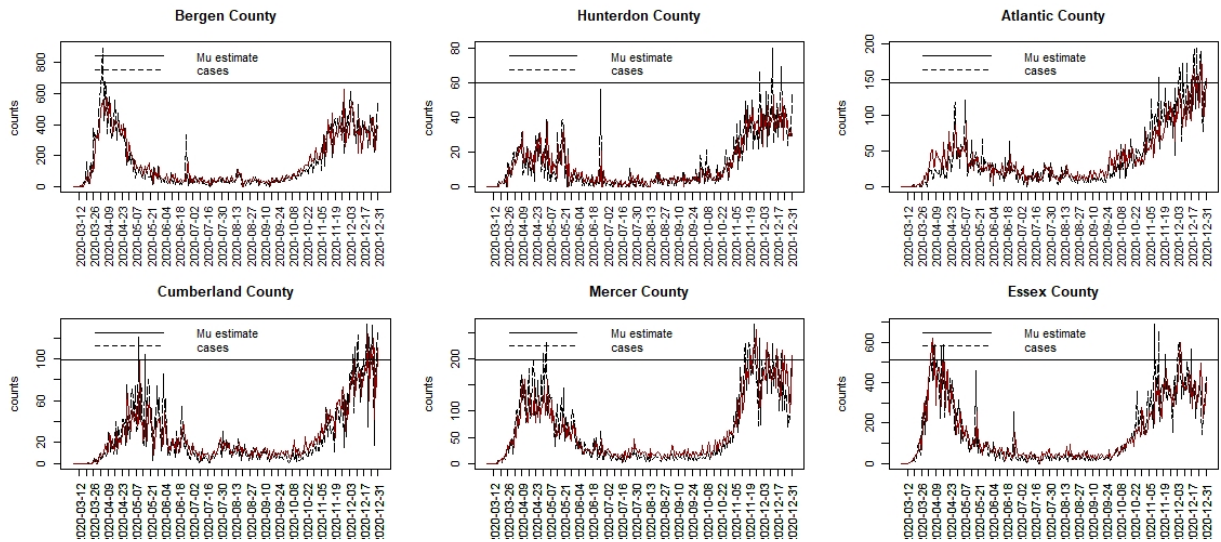